\documentclass[nofootinbib,prd,showpacs]{revtex4}
\usepackage{amsmath}
\usepackage{subfigure}
\usepackage{amsfonts}
\usepackage{hyperref}
\usepackage{amssymb}
\usepackage{graphicx}
\graphicspath{ {figs/} }

\usepackage[latin1]{inputenc}
\usepackage[english]{babel}
\usepackage{color,xcolor}

\definecolor{purple}{rgb}{1,0,1}
\definecolor{lime}{HTML}{A6CE39} 

\definecolor{lime}{HTML}{A6CE39}

\begin{document}
\title{
Orbits around a black bounce spacetime}

\author{Marcos V. de S. Silva}
\email{marco2s303@gmail.com}
\affiliation{Departamento de F\'isica, Programa de P\'os-Gradua\c c\~ao em F\'isica, Universidade Federal do Cear\'a, Campus Pici, 60440-900, Fortaleza, Cear\'a, Brazil}
\affiliation{Faculdade de F\'{\i}sica, Universidade Federal do Par\'{a},
 68721-000, Salin\'opolis, Par\'{a}, Brazil}
 
\author{Manuel E. Rodrigues}
\email{esialg@gmail.com}
\affiliation{Faculdade de Ci\^{e}ncias Exatas e Tecnologia, 
Universidade Federal do Par\'{a}\\
Campus Universit\'{a}rio de Abaetetuba, 68440-000, Abaetetuba, Par\'{a}, 
Brazil}
\affiliation{Faculdade de F\'{\i}sica, Programa de P\'{o}s-Gradua\c{c}\~ao em 
F\'isica, Universidade Federal do 
 Par\'{a}, 66075-110, Bel\'{e}m, Par\'{a}, Brazil}

\date{today; \LaTeX-ed \today}
\begin{abstract}
In this work, the trajectories of particles around a black bounce spacetime are considered, with the Simpson-Visser model serving as an example. Trajectories for massless and massive particles are obtained through the study of null and time-like geodesics. As the Simpson-Visser solution is derived via the Einstein equations for a source involving nonlinear electrodynamics and a scalar field, photon trajectories are investigated by considering an effective metric in which photons follow null geodesics. The stability of circular orbits is analyzed by examining the behavior of maxima and minima of the effective potential associated with geodesics. It is also studied what type of geodesic photons follow when the usual metric is considered instead of the effective one. The main focus of this work is to obtain corrections to the trajectories of photons when considering that the solution arises from nonlinear electrodynamics.
\end{abstract}
\maketitle
\def\HMS{{\scriptscriptstyle{HMS}}}
\section{Introduction}\label{S:intro}

In 1916, Karl Schwarzschild obtained the first solution to Einstein's equations, which describes the spacetime outside a spherically symmetric, static, and uncharged object \cite{Chandrasekhar:1985kt}. Initially, this solution represented only the exterior of a star, with the interior part considered non-physical for many years. This perspective persisted until it was recognized that the region inside the event horizon, the surface from which neither particles nor light can escape \cite{Wald:1984rg}, is indeed a part of the physical solution, containing a curvature singularity at its center. This solution was later known as a black hole \cite{Herdeiro:2018ldf}. 

Other black hole solutions have emerged over the years, such as the Reissner-Nordstrom solution, which has mass and charge as characteristic parameters, and the Kerr one, which has mass and angular momentum \cite{Chandrasekhar:1985kt, Visser:2007fj}. Several other black hole solutions appear in the literature, considering sources such as scalar field, spinorial field, and some other coupling with matter \cite{Bronnikov:2018vbs, Bronnikov:2022ofk}.

The spacetime of these solutions has some essential characteristics, the main one, which characterizes a black hole, is the causal structure \cite{Wald:1984rg}. To obtain the minimum length between two points in the spacetime, one must extremize the line element, that is, the quantity $s=\int \sqrt{g_{\mu\nu}\dot{x}^{\mu }\dot{x}^{\nu}}d\lambda$, where $g_{\mu\nu}$ is the metric, $\dot{x}^{\mu}=dx^{\mu}/ d\lambda$ with $\lambda$ being the affine parameter. These extreme curves are called geodesics, which can be space-like, time-like, and null \cite{Chandrasekhar:1985kt}. The causal structure of spacetime clearly shows how these geodesics are distributed in each considered region. For example, the event horizon is a null hypersurface, in which the outer part is causally separated from the inner part, with no time-like and null geodesics that go beyond this horizon from the inside out \cite{Wald:1984rg}. Massive particles follow time-like geodesics, and massless particles follow null geodesics \cite{Chandrasekhar:1985kt}.

With this approach, it becomes possible to characterize the type of black hole solution being described based on the optical properties of geodesics \cite{Guerrero:2021pxt, Guerrero:2021ues, Olmo:2021piq, Guerrero:2022qkh, Guerrero:2022msp, Olmo:2023lil, Rosa:2023qcv}. If the black hole is neutral, massive charged particles follow a particular path, which for a charged black hole would be a completely different one. If the black hole has rotation, the particles will follow a different trajectory from the static black hole. Consequently, by examining geodesics, various other black hole solutions can be identified  \cite{Chandrasekhar:1985kt}.

Recently, after many years of study, interactions of gravitational waves have been observed by the experiments of the LIGO/VIRGO collaboration, which is evidence of collisions of black holes in binary systems. 
 Another relevant experiment realized in recent years is the observation of a black hole shadow through the Event Horizon Telescope \cite{EventHorizonTelescope:2019dse,EventHorizonTelescope:2022wkp}. This image is strong evidence of how light follows null geodesics, forming a shadow characteristic of each black hole. The black hole's shadow occurs due to the bending of light passing around the black hole, creating an image seen by an observer very far away \cite{Cunha:2018acu}.

As the singularity of black holes is a point, or a set of points, in which Physics is not valid, other solutions were proposed in which this difficulty does not exist, regular solutions \cite{Ansoldi:2008jw}. The first nonsingular solution presented was the Bardeen solution \cite{Bardeen}. This solution has an event horizon but no singularity at any point in spacetime. This type of solution is known as a regular Bardeen black hole. After many years, Ayon-Beato and Garcia proposed that this solution came from a material content of nonlinear electrodynamics, magnetically charged \cite{Ayon-Beato:2000mjt}. After a few years, it is also shown that it can come from an electrically charged solution \cite{Rodrigues:2018bdc}. Several other regular black hole solutions have emerged over the years \cite{Ayon-Beato:1998hmi,Ayon-Beato:1999kuh,Bronnikov:2000yz,Dymnikova:2004zc,Balart:2009et,Culetu:2014lca,Rodrigues:2017yry,Rodrigues:2020pem,Simpson:2019mud,Bambi:2013ufa,Neves:2014aba,Toshmatov:2017zpr,Rodrigues:2017tfm,Franzin:2022wai,Singh:2022dqs,Torres:2022twv,Kubiznak:2022vft,Junior:2015fya,Rodrigues:2016fym,deSousaSilva:2018kkt,Rodrigues:2019xrc,Junior:2020zdt,Bronnikov:2005gm,Bronnikov:2006fu,Bronnikov:2012ch,Rodrigues:2022qdp}.

Other exact solutions of the Einstein equations are wormholes \cite{Visser:1995cc}. In particular, these solutions are regular and do not have horizons. They began to emerge with the Ellis-Bronnikov wormhole in 1973 \cite{Bronnikov:1973fh,Ellis:1973yv}. After that, several other solutions appear in the literature \cite{Visser:1989kh,Sushkov:2005kj,Visser:2003yf,Hochberg:1998ha,Bronnikov:2002rn,Kanti:2011jz,Barcelo:2000zf}. These structures always present a kind of bridge between spacetime regions called throats. In general, the throat is at the center of the radial coordinate.

Recently, a solution that merges the idea of a regular black hole and a wormhole, with a throat in the center, is the Simpson-Visser spacetime, where there is a parameter in which, for a particular value, the solution describes a regular black hole with throat or a wormhole \cite{Simpson:2018tsi}. When the parameter is zero, the Schwarzschild solution is recovered. Some analyses are made in the literature \cite{Lima:2020auu,LimaJunior:2022zvu,Lobo:2020kxn,Tsukamoto:2020bjm,Bambhaniya:2021ugr}. Other solutions were proposed in the literature \cite{Lima:2022pvc,Furtado:2022tnb,Junior:2022zxo,Guo:2021wid,Franzin:2021vnj,Lobo:2020ffi,Huang:2019arj,Rodrigues:2022mdm,Akil:2022coa}. There are also solutions with a cloud of strings \cite{Rodrigues:2022rfj,Yang:2022ryf}. Interestingly, this solution cannot be modeled only by nonlinear electrodynamics or a scalar field. Still, it can be an exact solution of the equations arising from the coupling between nonlinear electrodynamics and a phantom scalar field, where now the bounce parameter is the magnetic charge \cite{Canate:2022gpy,Bronnikov:2021uta,Rodrigues:2023vtm,Bronnikov:2023aya}. Also, there are works in the literature that show that black bounce solutions can also be obtained through electric sources \cite{Lima:2023arg,Alencar:2024yvh}. Therefore, knowing that photons traveling in spacetime that have a nonlinear electrodynamics source do not follow null geodesics but rather an effective null geodesic \cite{NED4,Toshmatov:2021fgm}, in this work, the main characteristics of all geodesics for massive particles, massless particles, and photons will be examined.

Despite corrections from nonlinear electrodynamics having been well studied in the context of regular black holes, this kind of study has not been done so far in the context of black bounces.

The work is organized as follows: Section \ref{S:geo_geral} introduces the equations for time-like, null, and effective null geodesics. Section \ref{S:PO-SV} presents the Simpson-Visser solution and its associated material content, addressing null-type, time-type, and effective null-type geodesics. Finally, Section \ref{S:conclusion} provides the concluding remarks.

\section{Geodesics in black bounce spactimes}\label{S:geo_geral}
The line element that describes a black bounce spacetime can be written as
\begin{equation}
    ds^2=f(r)dt^2-\frac{1}{f(r)}dr^2-\Sigma(r)^2\left(d\theta^2+\sin^2\theta d\varphi^2\right).\label{Line}
\end{equation}
However, for reasons that will become clear shortly, the line element will be written as: 
\begin{equation}
    ds^2=A(r)dt^2-B(r)dr^2-C(r)\left(d\theta^2+\sin^2\theta d\varphi^2\right).\label{geoline}
\end{equation}

To obtain the trajectories of particles in these spacetimes, the Lagrangian is considered as follows
\begin{equation}
    \mathcal{L}=\dot{s}^2=A\dot{t}^2-B\dot{r}^2-C\left(\dot{\theta}^2+\sin^2\theta \dot{\varphi}^2\right),\label{geoLagran}
\end{equation}
where the dot represents the derivative with respect to the affine parameter $\tau$. Using the Euler-Lagrange equations
\begin{equation}
    \frac{d}{d\tau}\left(\frac{\partial \mathcal{L}}{\partial \dot{x^\mu}}\right)-\frac{\partial \mathcal{L}}{\partial x^\mu}=0,\label{Euler-Lagran}
\end{equation}
the equations of motion are
\begin{eqnarray}
    A\dot{t}=E,\label{geo1}\\
   A\dot{t}^2-B\dot{r}^2-C\left(\dot{\theta}^2+\sin^2\theta \dot{\varphi}^2\right)=\delta,\label{geo2}\\
   \ddot{\theta}+\frac{C'}{C}\dot{\theta}\dot{r}-\sin\theta\cos\theta \dot{\varphi}^2=0,\label{geo3}\\
   C\sin^2\theta \dot{\varphi}=\ell,\label{geo4}
\end{eqnarray}
where $E$ and $\ell$ are the energy and the angular moment of the particle \cite{Chandrasekhar:1985kt}. The constant $\delta$ assume the values $\delta=1$, for massive particles, and $\delta=0$, for massless particles. Since the spacetime is spherically symmetric, without any loss of generality, the equatorial plane can be chosen, $\theta=\pi/2$. With this, equation \eqref{geo3} is automatically satisfied, and the remaining equations are
\begin{eqnarray}
    A\dot{t}=E,\label{ggeo1}\\
   A\dot{t}^2-B\dot{r}^2-C\dot{\varphi}^2=\delta,\label{ggeo2}\\
    C\dot{\varphi}=\ell.\label{ggeo3}
\end{eqnarray}
\subsection{Geodesics of massless particles}
For massless particles $\delta=0$.
Substituting \eqref{ggeo1} and \eqref{ggeo3} in \eqref{ggeo2} yields
\begin{equation}
    E^2=AB\dot{r}^2+\frac{Al^2}{C}.
\end{equation}
This is a type of energy conservation equation. This equation can be reformulated as
\begin{equation}
    \dot{r}^2=\frac{1}{AB}\left(E^2-V_{eff}\right),\label{energicons}
\end{equation}
where $V_{eff}$ is the effective potential given by
\begin{equation}
    V_{eff}= \frac{Al^2}{C}.\label{PotGeneral}
\end{equation}

In the case of the black bounce solutions under consideration in this article, the metric coefficients can be expressed as $A(r)=B(r)^{-1}=f(r)$ and $C(r)=\Sigma(r)^2$. This simplification leads to
\begin{equation}
    \dot{r}^2=E^2-V_{eff},\label{EnerConsMassless}
\end{equation}
with
\begin{equation}
    V_{eff}= \frac{fl^2}{\Sigma^2}.
\end{equation}

\subsection{Effective metrics for photon orbits in a black bounce spacetime}\label{S:PO-BB}

According to \cite{Canate:2022gpy,Bronnikov:2021uta,Rodrigues:2023vtm}, some black bounce spacetimes can be considered solutions of Einstein equations for a phantom scalar field with nonlinear electrodynamics. However, in the presence of nonlinear electrodynamics, photons do not follow null geodesics. Instead, photons will follow null geodesics in an effective metric given by \cite{NED4,Toshmatov:2021fgm}
\begin{equation}
    g^{\mu\nu}_{eff}=L_F g^{\mu\nu}-L_{FF}{F^{\mu}}_{\lambda}F^{\lambda\nu}.\label{metriceff}
\end{equation}
The line element associated with this metric is
\begin{equation}
    ds^2=\frac{f}{L_F}dt^2-\frac{1}{L_F f}dr^2-\frac{\Sigma^2}{L_F+2FL_{FF}}\left(d\theta^2+\sin^2\theta d\varphi^2\right),\label{LineEff}
\end{equation}
where $L_F$ and $L_{FF}$ are the first and the second derivative of the electromagnetic Lagrangian, $L(F)$, with respect to the electromagnetic scalar $F=F^{\mu\nu}F_{\mu\nu}/4$, with $F^{\mu\nu}$ being the Maxwell-Faraday tensor.

For this situation, it is necessary to consider equations \eqref{energicons} and \eqref{PotGeneral} with
\begin{equation}
    A=\frac{f}{L_F}, \quad B= \frac{1}{L_F f}, \quad \mbox{and} \quad C=\frac{\Sigma^2}{L_F+2FL_{FF}}.\label{effcompo}
\end{equation}

\subsection{Geodesics of massive particles}
For massive particles, $\delta=1$, and then
\begin{equation}
    f\dot{t}^2-f^{-1}\dot{r}^2-\Sigma^2\dot{\varphi}^2=1.
\end{equation}
Using \eqref{ggeo1} and \eqref{ggeo3}, it is obtained that
\begin{equation}
    E^2=\dot{r}^2+V_{eff},\label{EnerConsMassive}
\end{equation}
where the effective potential is
\begin{equation}
    V_{eff}=f\left(1+\frac{l^2}{\Sigma^2}\right).
\end{equation}
The next step in the study of geodesics is to specify the solutions to be analyzed.

\section{Simpson-Visser spacetime}\label{S:PO-SV}
The Simpson-Visser solution arises as a regularization proposal for the Schwarzschild solution, where the radial coordinate $r$ is replaced by $r\rightarrow \sqrt{a^2 +r^2}$, with $a$ being the parameter responsible for spacetime regularization. The line element that describes this spacetime is written as \cite{Simpson:2018tsi}
\begin{equation}
    ds^2=\left(1-\frac{2m}{\sqrt{r^2+a^2}}\right)dt^2-\left(1-\frac{2m}{\sqrt{r^2+a^2}}\right)^{-1}dr^2-\left(r^2+a^2\right)\left(d\theta^2+\sin^2\theta d\varphi^2\right).\label{SV-model}
\end{equation}
Depending on the value of the regularization parameter, the following scenarios can be observed: when $a > 2m$, a two-way traversable wormhole exists; for $a = 2m$, a one-way traversable wormhole is present; when $a < 2m$, a regular black hole with a localized throat inside is observed; and when $a = 0$, the Schwarzschild solution is recovered. Since this spacetime is regular, all curvature invariants do not exhibit divergences, while also violating the null energy condition \cite{Simpson:2018tsi}.

The Simpson-Visser solution can also be obtained from the Einstein equations for a source which is described  by the coupling of a nonlinear electrodynamics with a phantom scalar field. These fields are described by
\begin{eqnarray}
   \phi(r)&=&\frac{\tan ^{-1}\left(\frac{r}{q}\right)}{\kappa},\quad
    V(\phi)=\frac{4 m \cos ^5\left(\phi 
   \kappa \right)}{5 \kappa ^2 \left|q\right|^{3}},\\
   L(F)&=&\frac{12 \sqrt[4]{2} m F^{5/4} }{5 \kappa ^2
   \sqrt{\left|q\right|}},\quad
   F(r)=\frac{q^2}{2\Sigma^4},\label{F}
\end{eqnarray}
where $V(\phi)$ is the potential related to the scalar field $\phi$, and $\kappa=8\pi$. To obtain this source, it was considered that the regularization parameter is the charge of the black hole, $a\rightarrow q$.
\subsection{Massless particles}
\begin{figure}
    \centering
    \includegraphics[scale=0.7]{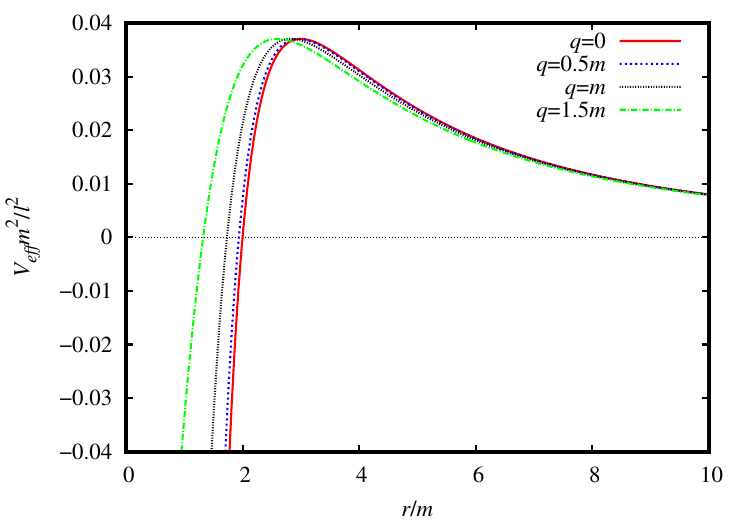}
    \includegraphics[scale=0.7]{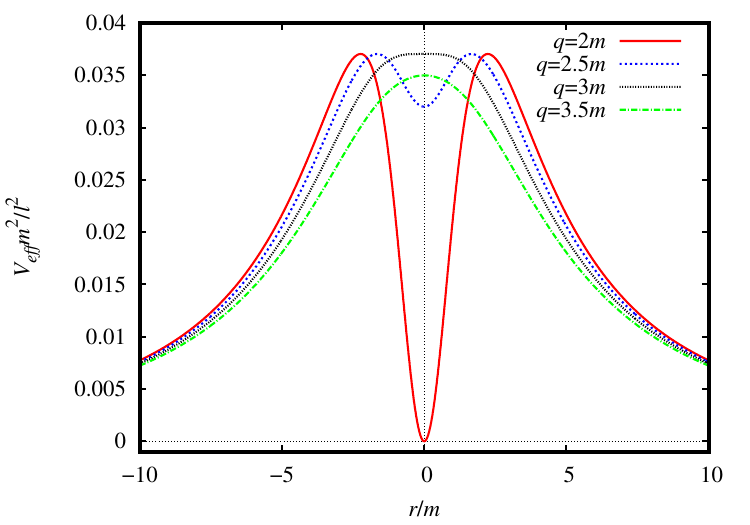}
    \caption{Effective potential for null geodesics in the Simpson-Visser spacetime with different values of charge.}
    \label{Fig-Veff-SV-Massless}
\end{figure}

The effective potential for massless particles is given by
\begin{equation}
V_{eff}= \frac{l^2 \left(\sqrt{q^2+r^2}-2 m\right)}{\left(q^2+r^2\right)^{3/2}}.\label{Veffmassless}
\end{equation}
The potential goes to zero at infinity and at the event horizon. Depending on the charge value, there are three possible situations: for $q<2m$ there is only one maximum in the effective potential; for $2m\leq q < 3m$ there are two maximums and one maximum; for $q\geq3m$ there is only one minimum. To obtain the extreme points, it is necessary to solve $V'(r)=0$, that results in
\begin{equation}
    r_{1}= \sqrt{9m^2-q^2}, \quad r_{2}=- \sqrt{9m^2-q^2},\quad \mbox{and} \quad r_{3}=0.\label{extremeSV}
\end{equation}
In Fig. \ref{Fig-Veff-SV-Massless}, the effective potential for null geodesics is depicted. For $q<2m$, the maximum is situated at $r=r_{1}$, representing the radius of an unstable circular orbit for the particle. For $2m\leq q < 3m$, the maxima are located at $r=r_{1}$ and $r=r_{2}$, while the minimum is at $r=r_{3}$. Consequently, there exist two possible unstable orbits and one stable massless particle orbit. Finally, for $q\geq 3m$, only one unstable orbit exists at $r=r_{3}$.

\begin{figure}
    \centering
    \includegraphics[scale=0.7]{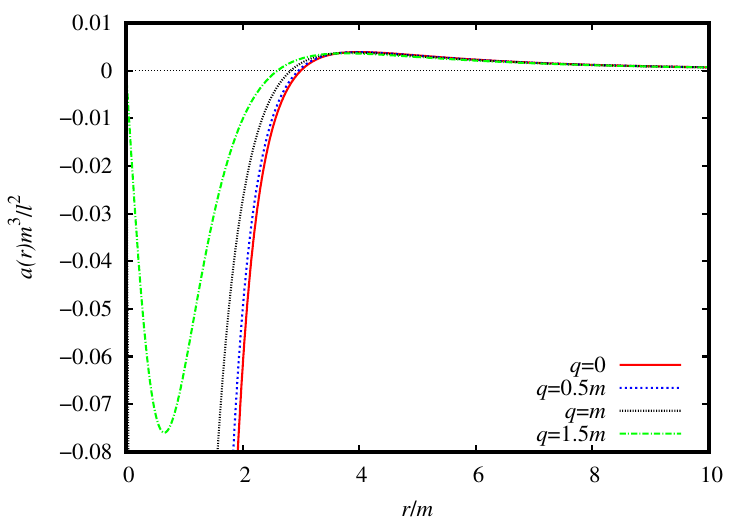}
    \includegraphics[scale=0.7]{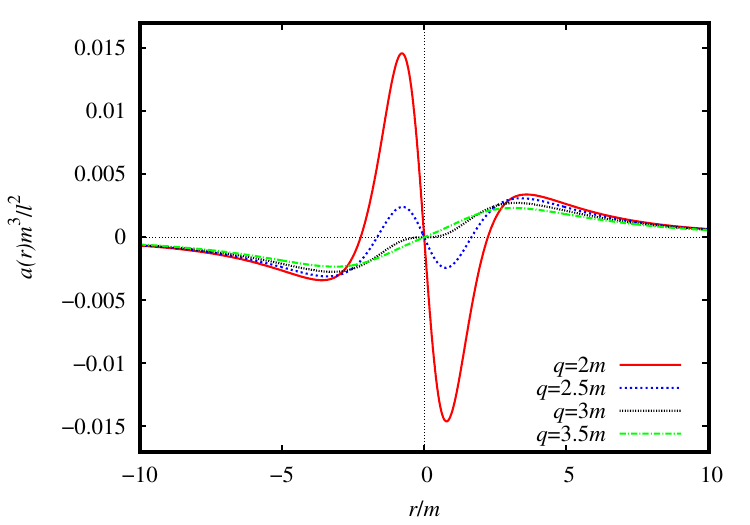}
    \caption{Radial acceleration for massless particles in the Simpson-Visser spacetime to different values of charge.}
    \label{Fig-a-SV-Massless}
\end{figure}

The radial acceleration of the massless particle can be obtained as
\begin{equation}
    a=\ddot{r}=-\frac{1}{2}\frac{dV_{eff}}{dr}=\frac{l^2 r \left(\sqrt{q^2+r^2}-3
   m\right)}{\left(q^2+r^2\right)^{5/2}}.\label{a_SV_massless}
\end{equation}
When the radial acceleration is computed at the extremum points of the effective potential, the result is $a(r_1,r_2,r_3)=0$. Notably, the radial acceleration exhibits asymmetry with respect to $r\rightarrow -r$. In Fig. \ref{Fig-a-SV-Massless}, the radial acceleration for massless particles in the Simpson-Visser spacetime is presented. It is observed that when $q<2m$, the radial acceleration is positive (repulsive) for $r>r_1$ and negative (attractive) for $r<r_1$. For $2m\leq q < 3m$, the radial acceleration is negative within the interval $r_3<r<r_1$ and $r<r_2$, while it is positive for $r>r_1$ and $r_2<r<r_3$.

Consider a particle that comes from infinity with energy $E_C^2$ that is equal to the maximum of the effective potential. In this case, when the particle reaches a maximum point of the effective potential, both the radial velocity and the radial acceleration are zero, $\dot{r}=\ddot{r}=0$. From equations \eqref{ggeo1} and \eqref{ggeo2}, it can be find that
\begin{equation}
    \dot{\varphi}^2=\pm \frac{E_c}{f(r_{1,2,3})^{1/2}(r_{1,2,3}^2+q^2)^{1/2}}.
\end{equation}
It means that, despite the radial velocity be zero, the angular velocity is nonzero. The particle describes a circular orbit.

The critical impact parameter is defined as \cite{Chandrasekhar:1985kt}
\begin{equation}
    b_c=\frac{l_c}{E_c}=(r_c^2+q^2)^{1/2}\left(1-\frac{2m}{\sqrt{q^2+r_c^2}}\right)^{-1/2},
\end{equation}
where $l_c$ and $E_c$ are parameters related to the circular orbit. Here, $r_c=r_1$ for $0\leq q< 3m$ and $r_c=r_3$ if $q\geq 3m$. Substituting $r_1$ and $r_3$, the following expressions are obtained
\begin{eqnarray}
    b_c&=&3\sqrt{3}m, \quad \mbox{for} \quad \left|q\right|< 3m,\\
    b_c&=&\frac{\left|q\right|^{3/2}}{\sqrt{\left|q\right|-2m}}, \quad \mbox{for} \quad \left|q\right|\geq 3m.
\end{eqnarray}
For $\left|q\right|<3m$, the critical impact parameter is the same of the Schwarzschild case. The charge affects the critical impact parameter only for $\left|q\right|\geq3m$. If a particle has an impact parameter greater than the critical one, it will be scattered. If the particle has an impact parameter smaller than the critical one, it will be absorbed. Finally, if the impact parameter is equal to the critical one then there is a circular orbit. Based on this, the critical impact parameter can be used to calculate the absorption cross section for massless particles, $\sigma=\pi b_c^2$, which is
\begin{eqnarray}
    \sigma&=&27\pi m^2, \quad \mbox{for} \quad \left|q\right|< 3m,\\
    \sigma&=&\frac{\pi \left|q\right|^{3}}{\left|q\right|-2m}, \quad \mbox{for} \quad \left|q\right|\geq 3m.
\end{eqnarray}
For $\left|q\right|<3m$, the absorption cross section is the same of the Schwarzschild solution. For $\left|q\right|>3m$ the absorption cross section will grow with the charge.

Making the change of variable
\begin{equation}
    u=\frac{1}{\Sigma}=\frac{1}{\sqrt{r^2+q^2}},
\end{equation}
and substituting in the equation \eqref{EnerConsMassless}, results in
\begin{equation}
    \left(\frac{du}{d\varphi}\right)^2=\left(1-q^2u^2\right)\left(\frac{1}{b^2}+u^2\left(2mu-1\right)\right).\label{du}
\end{equation}

Taking the derivative of equation \eqref{du} with respect to $\varphi$ yields
\begin{equation}
    \frac{d^2u}{d\varphi^2}+u+\frac{q^2 u}{b^2}+5 m q^2 u^4-3 m u^2-2 q^2 u^3=0.\label{EQ2Geo-SV-massless}
\end{equation}
To solve the equation \eqref{EQ2Geo-SV-massless}, it is necessary two boundary conditions. For $r\rightarrow \infty$, both $\varphi$ and $u$ tend to zero, $u(\phi=0)=0$. Our second boundary condition can be obtained from equation \eqref{du} as $r$ tends to infinity, $u$ tends to zero, resulting in:
\begin{equation}
    \left.\frac{du}{d\varphi}\right|_{\varphi =0}=\pm\frac{1}{b}.\label{du2}
\end{equation}

The condition that $\varphi=0$ at spatial infinity means that the geodesic comes from future infinity, from a horizontal direction, at a distance $b$ (impact parameter) from the $x$ axis, because the angle $\varphi$ is the one that the radial polar direction $r$ makes with the horizontal axis $x$.

\begin{figure}
    \centering
   \subfigure[]{\includegraphics[scale=0.8]{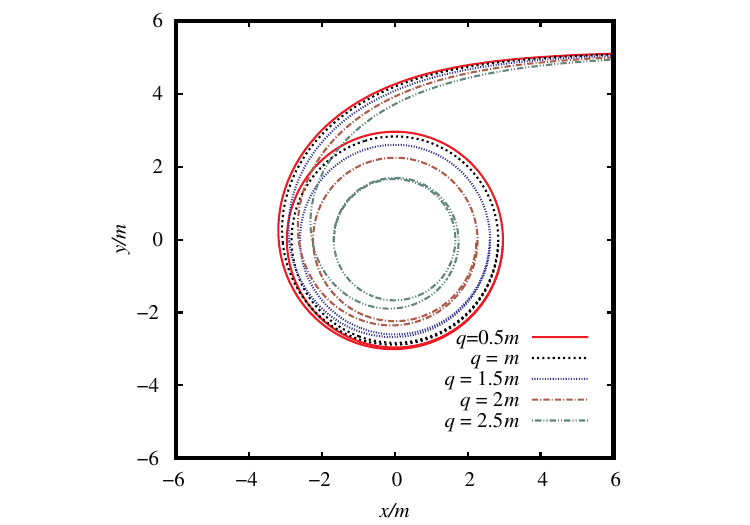}}\hspace{-3cm}
    \subfigure[]{\includegraphics[scale=0.8]{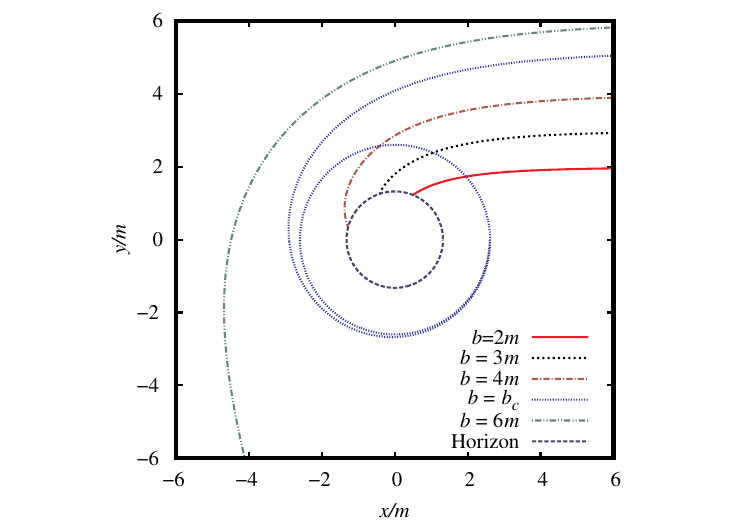}}
    \caption{Trajectories of massless particles in the Simpson-Visser spacetime are examined. In panel (a), the impact parameter is held constant at $b=b_c=3\sqrt{3}m$, while the magnetic charge is varied. In panel (b), the charge is fixed at $q=1.5m$, and the impact parameter is altered.}
    \label{fig:SV-geo-massless}
\end{figure}

Once the boundary conditions are established, the geodesic curves can be generated. As shown in Fig. \ref{fig:SV-geo-massless}, it becomes apparent that with an increase in charge, the radius of the photon orbit decreases. Additionally, it is observed that particles with an impact parameter smaller than the critical value are absorbed by the black hole, while those with an impact parameter exceeding the critical value are scattered.

The results present in this subsection are consistent with \cite{Lima:2020auu,LimaJunior:2022zvu}.
\subsection{Photons}
Considering \eqref{effcompo}, \eqref{SV-model}, and \eqref{F}, the effective potential is
\begin{equation}
    V_{eff}=\frac{3 l^2 \left(\sqrt{q^2+r^2}-2 m\right)}{2 \left(q^2+r^2\right)^{3/2}}.\label{Veff-SV}
\end{equation}

\begin{figure}
    \centering
    \includegraphics[scale=0.5]{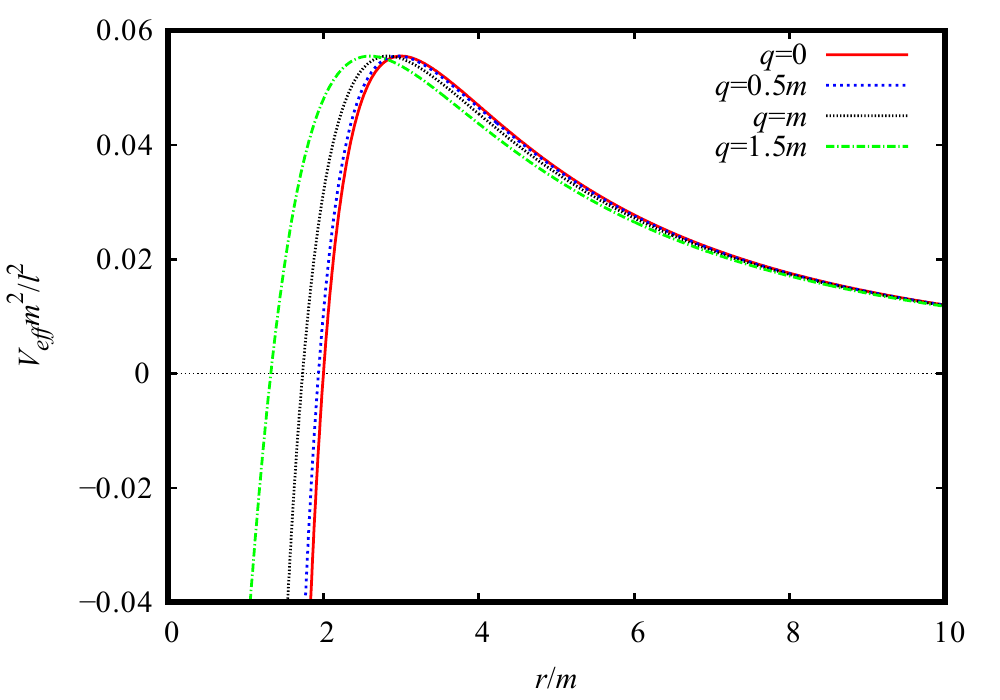}\hspace{-.4cm}
    \includegraphics[scale=0.5]{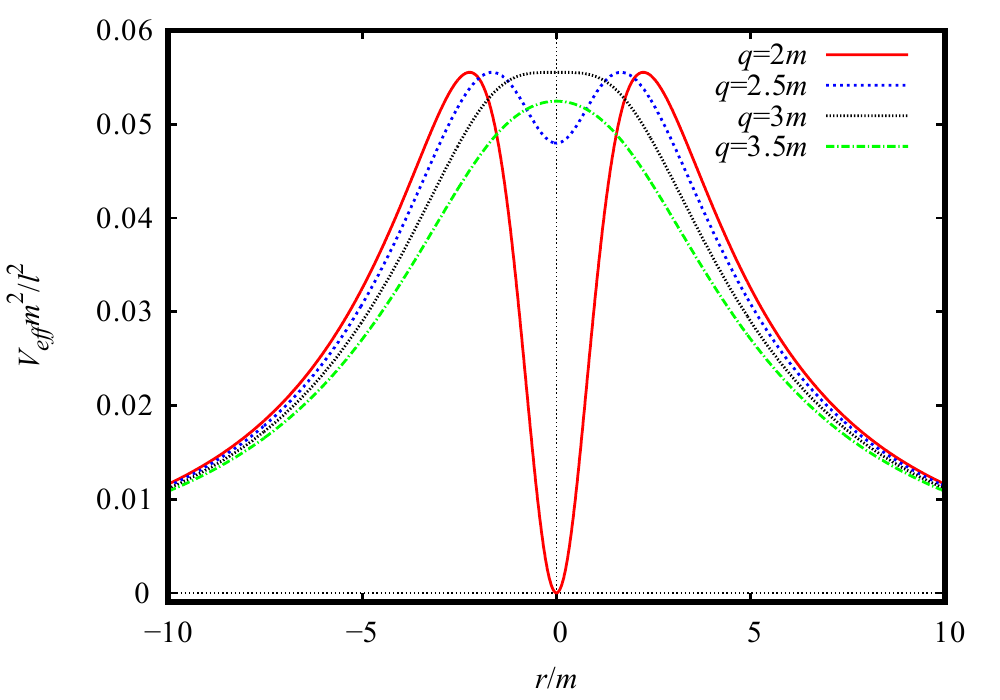}
    \caption{Effective potential for photons in the effective line element that describes the Simpson-Visser model for different values of charge.}
    \label{Fig-Veff-SV}
\end{figure}

In Fig. \ref{Fig-Veff-SV}, the effective potential for photons is presented, showcasing variations for different charge values. The behavior of the effective potential closely resembles that of null geodesics. However, it's worth noting that the effective potential for photons exhibits higher values compared to other massless particles. This implies that photons must possess greater energy to be absorbed by the black bounce. The determination of extreme points involves solving $V_{eff}'(r)=0$, which yields the same values as \eqref{extremeSV}. Despite accounting corrections arising from the nonlinear electrodynamics, the location of the photon orbit remains consistent with the case where these corrections are not considered.


The critical impact parameter is
\begin{equation}
    b_c=\frac{2}{3}(r_c^2+q^2)^{1/2}\left(1-\frac{2m}{\sqrt{q^2+r_c^2}}\right)^{-1/2}.
\end{equation}
Substituting $r_1$ and $r_3$, which remain the same from the massless case, is obtained
\begin{eqnarray}
    b_c&=&3\sqrt{2}m, \quad \mbox{for} \quad \left|q\right|< 3m,\\
    b_c&=&\sqrt{\frac{2}{3}}\frac{\left|q\right|^{3/2}}{\sqrt{\left|q\right|-2m}}, \quad \mbox{for} \quad \left|q\right|\geq 3m.
\end{eqnarray}
Although the behavior is very similar to the case where there are nonlinear electrodynamics corrections, the value of the critical impact parameter changes due to these corrections. There are also modifications on the absorption cross section, that are given by
\begin{eqnarray}
    \sigma&=&18\pi m^2, \quad \mbox{for} \quad \left|q\right|< 3m,\\
    \sigma&=&\frac{2}{3}\frac{\pi \left|q\right|^{3}}{\left|q\right|-2m}, \quad \mbox{for} \quad \left|q\right|\geq 3m.
\end{eqnarray}
The absorption cross section will be smaller for the case of photons than for massless particles. This is connected to the fact that the effective potential is greater for photons and therefore they need to be more energetic to be absorbed.

Using the change of variable
\begin{equation}
    u=\frac{1}{\Sigma}=\frac{1}{\sqrt{r^2+q^2}},
\end{equation}
and substituting in the equation \eqref{energicons}, the result is
\begin{equation}
    \left(\frac{du}{d\varphi}\right)^2=\frac{2\left(1-q^2u^2\right)}{9}\left(\frac{2}{b^2}+3u^2\left(2mu-1\right)\right).\label{dumass}
\end{equation}

\begin{figure}
    \centering
   \subfigure[]{\includegraphics[scale=0.6]{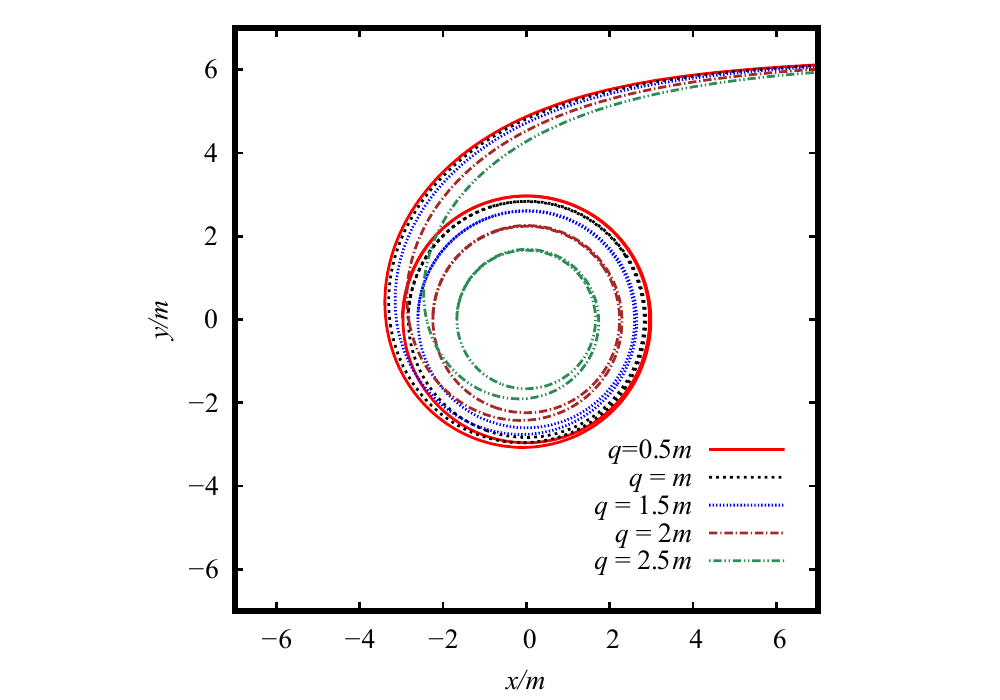}}\hspace{-3cm}
    \subfigure[]{\includegraphics[scale=0.8]{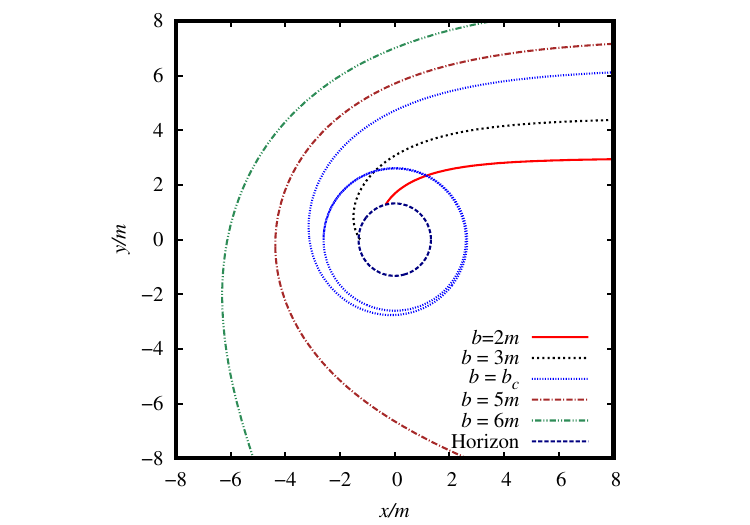}}
    \caption{Photon trajectories within the Simpson-Visser spacetime, accounting nonlinear electrodynamics corrections. In (a) there is a fixed value of the impact parameter, $b=b_c=3\sqrt{2}m$, with varying magnetic charge. In (b), the charge is held constant at $q=1.5m$, while the impact parameter value is altered.}
    \label{fig:SV-geo}
\end{figure}

The derivation of the equation \eqref{dumass} with respect to $\varphi$ results in
\begin{equation}
    \frac{d^2u}{d\varphi^2}+\frac{2 u}{3}+\frac{4 q^2 u}{9 b^2}-2 m u^2-\frac{4 q^2
   u^3}{3}+\frac{10}{3} m q^2 u^4=0.\label{EQ2Geo-SV}
\end{equation}
To solve the equation \eqref{EQ2Geo-SV}, two boundary conditions are required. As the case before, for $r\rightarrow \infty$ both $\varphi$ and $u$ tend to zero, $u(\phi=0)=0$ (see the explanation below of the equation \eqref{du2}). Imposing $u\rightarrow 0$ ($\phi \rightarrow 0$, $r\rightarrow \infty$) in equation \eqref{dumass} gives the second boundary condition, which is
\begin{equation}
    \left.\frac{du}{d\varphi}\right|_{\varphi =0}=\pm\frac{2}{3b}.
\end{equation}
In Fig. \ref{fig:SV-geo}, it is observed that with an increase in charge, the radius of the photon orbit decreases. Furthermore, it can be noted that particles with an impact parameter smaller than the critical impact parameter are absorbed by the black hole, while particles with an impact parameter greater than the critical impact parameter are scattered.

\subsection{Massive particles}
\begin{figure}
    \centering
    \subfigure[]{\includegraphics[scale=0.55]{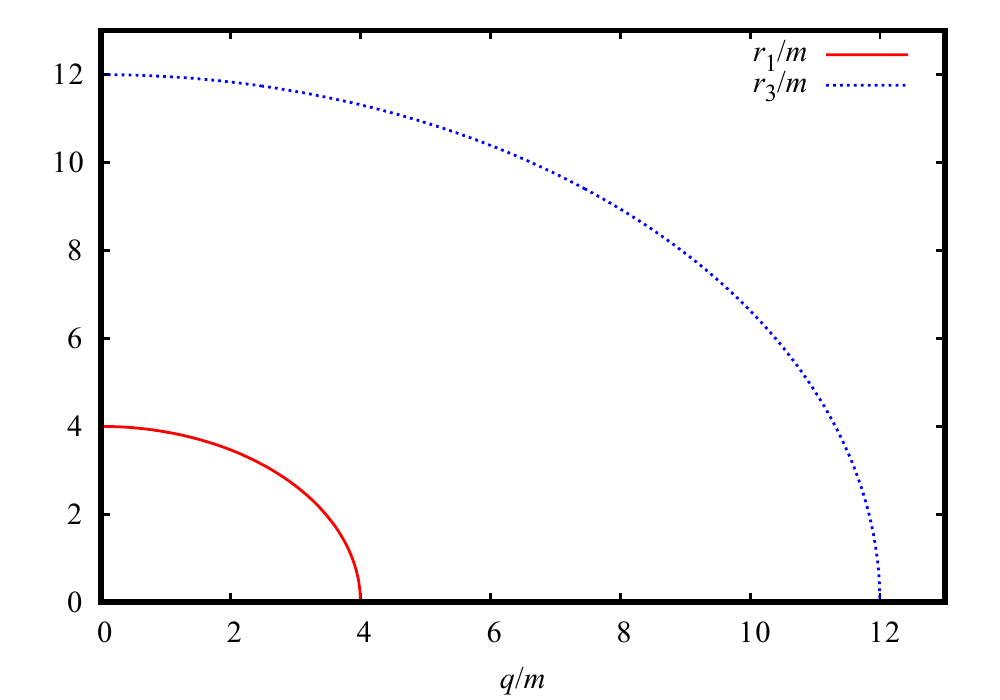}}\hspace{-.8cm}
    \subfigure[]{\includegraphics[scale=0.55]{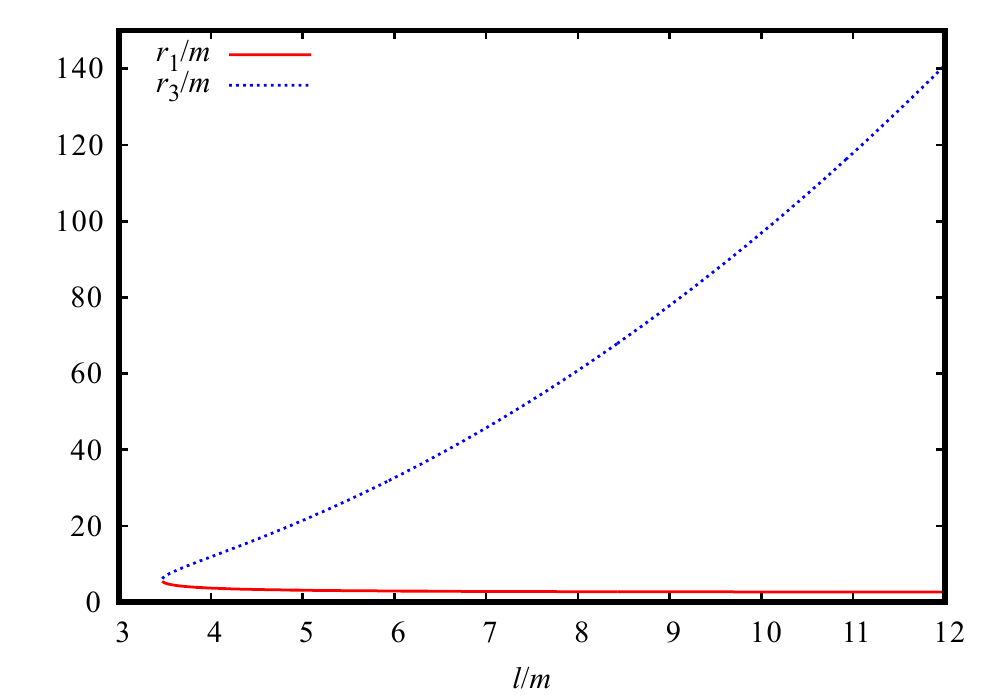}}
    \caption{Radius of the stable and the unstable orbit for massive particles in the Simpson-Visser spacetime time as functions of the charge and $l$. In (a) it is consider $l=4m$, and in (b) $q=1.5m$.}
    \label{fig:SV-radiusmassive}
\end{figure}

For massive particles, the effective potential is
\begin{figure}
    \centering
    \subfigure[]{\includegraphics[scale=0.5]{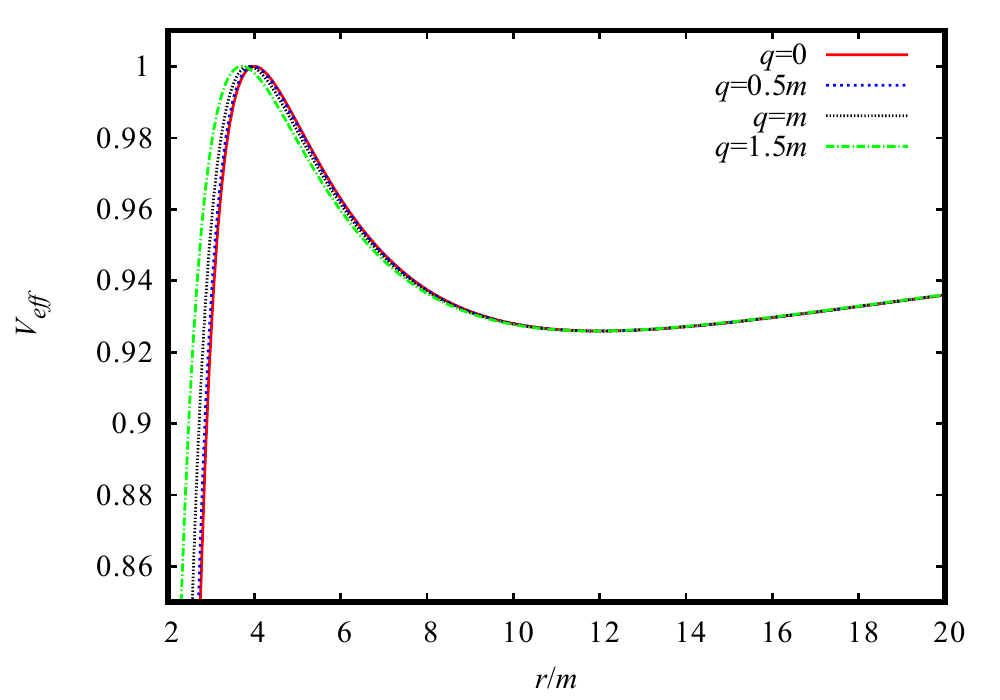}}\hspace{-.5cm}
    \subfigure[]{\includegraphics[scale=0.5]{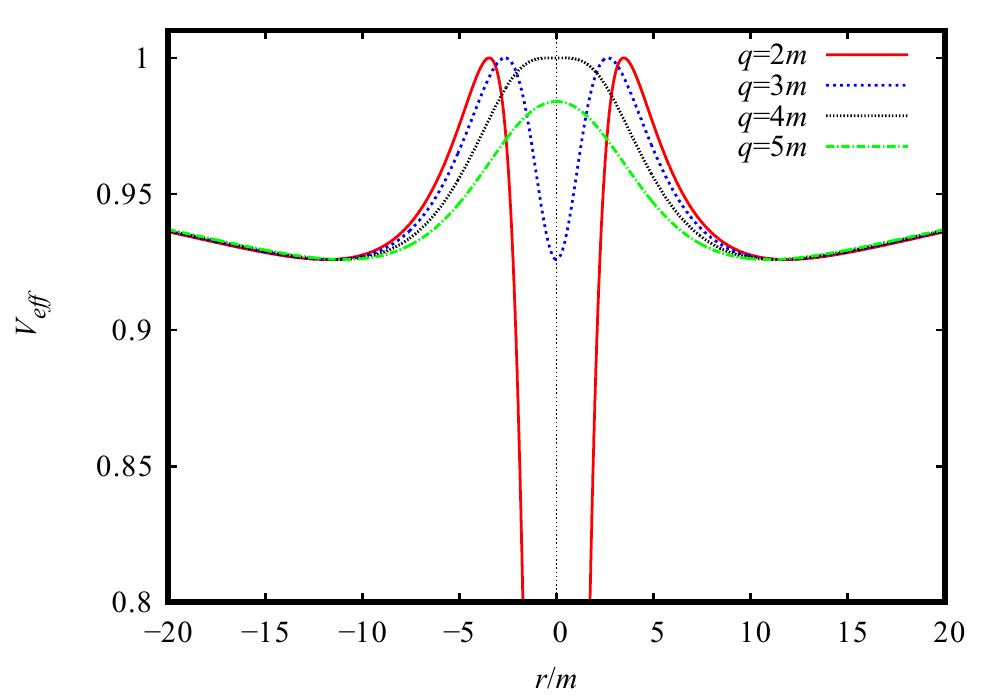}}\hspace{-.5cm}
    \subfigure[]{\includegraphics[scale=0.5]{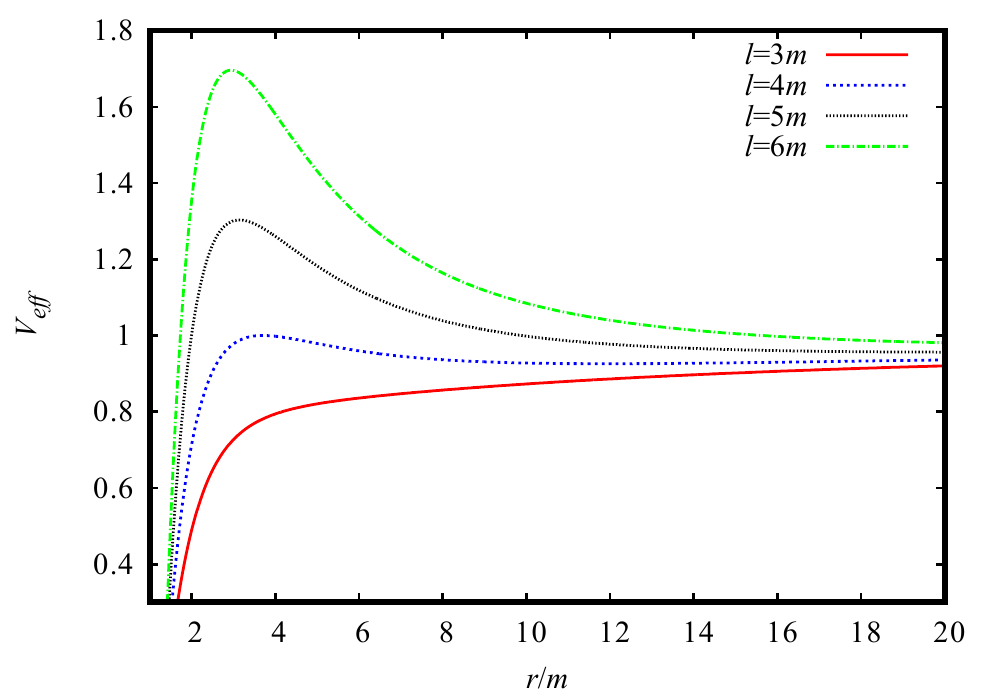}}\hspace{-.5cm}
    \subfigure[]{\includegraphics[scale=0.5]{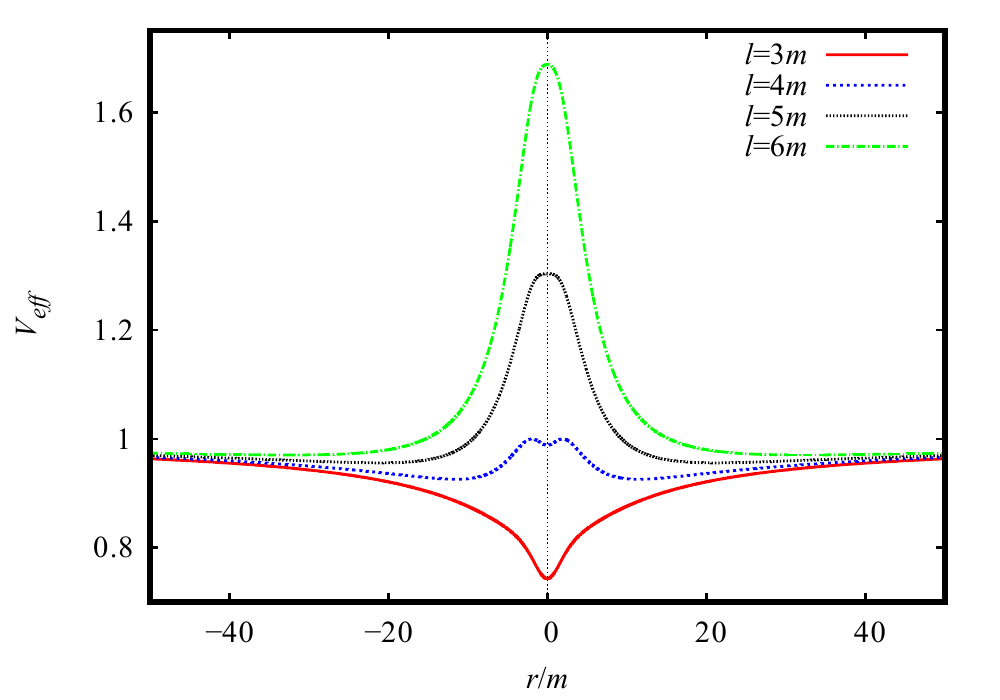}}
    \caption{The effective potential for massive particles in the Simpson-Visser spacetime is examined with varying values of charge and $l$. In (a), $l$ is held constant at $l=4m$, and the charge is adjusted. Given that $q<2m$, an event horizon exists, only the region $r>0$ is considered. In (b), the setup is similar to (a), but in this case, there is no event horizon, and the entire range $-\infty<r<\infty$ is taken into account. In (c) and (d), the charge is fixed at $q=1.5m$ and $q=3.5m$, respectively, while $l$ is varied.}
    \label{Fig-Veff-SV-Massive}
\end{figure}
\begin{equation}
    V_{eff}=\left(1+\frac{l^2}{q^2+r^2}\right) \left(1-\frac{2
   m}{\sqrt{q^2+r^2}}\right).\label{Veff_massive_SV}
\end{equation}
In Fig. \ref{Fig-Veff-SV-Massive} the behavior of the effective potential for different values of $q$ and $l$ is presented. The effective potential can have up to five extremum points. Depending on the values of the parameters, there are more maximum or minimum points. As the charge increases, the extreme points tend to $r=0$. For small values of $l$ there is no maximum points, it means that there is no unstable orbit. Regardless of the energy of the particle, if it came from infinity, it will not be scattered. Solving the condition $V'_{eff}=0$, the extremum points are
\begin{eqnarray}
    r_1&=& \frac{\sqrt{\frac{l^4}{m^2}-\frac{\sqrt{l^8-12 l^6 m^2}}{m^2}-6 l^2-2
   q^2}}{\sqrt{2}}, \quad r_2=-\frac{\sqrt{\frac{l^4}{m^2}-\frac{\sqrt{l^8-12 l^6 m^2}}{m^2}-6 l^2-2
   q^2}}{\sqrt{2}},\nonumber\\
   r_3&=&\frac{\sqrt{\frac{l^4}{m^2}+\frac{\sqrt{l^8-12 l^6 m^2}}{m^2}-6 l^2-2
   q^2}}{\sqrt{2}}\quad r_4=-\frac{\sqrt{\frac{l^4}{m^2}+\frac{\sqrt{l^8-12 l^6 m^2}}{m^2}-6 l^2-2
   q^2}}{\sqrt{2}}, \quad r_5=0.
\end{eqnarray}
As $r_1=-r_2$ and $r_3=-r_4$, only positive radii are considered. In Fig. \ref{fig:SV-radiusmassive}, it is observed that the radius decreases as the charge increases. Beyond a certain charge threshold, all radii tend to converge to $r_5$. There exists a minimum value of $l$ for the radius to exist. 

In Schwarzschild spacetime, any massive particles are absorbed by the black hole when $l^2<12m^2$, and for $l^2=12m^2$, the smallest stable orbit is established. In fact, at this value, the stable and unstable orbits converge to the same position, known as the innermost stable circular orbit (ISCO).

\begin{figure}
    \centering
    \includegraphics[scale=0.55]{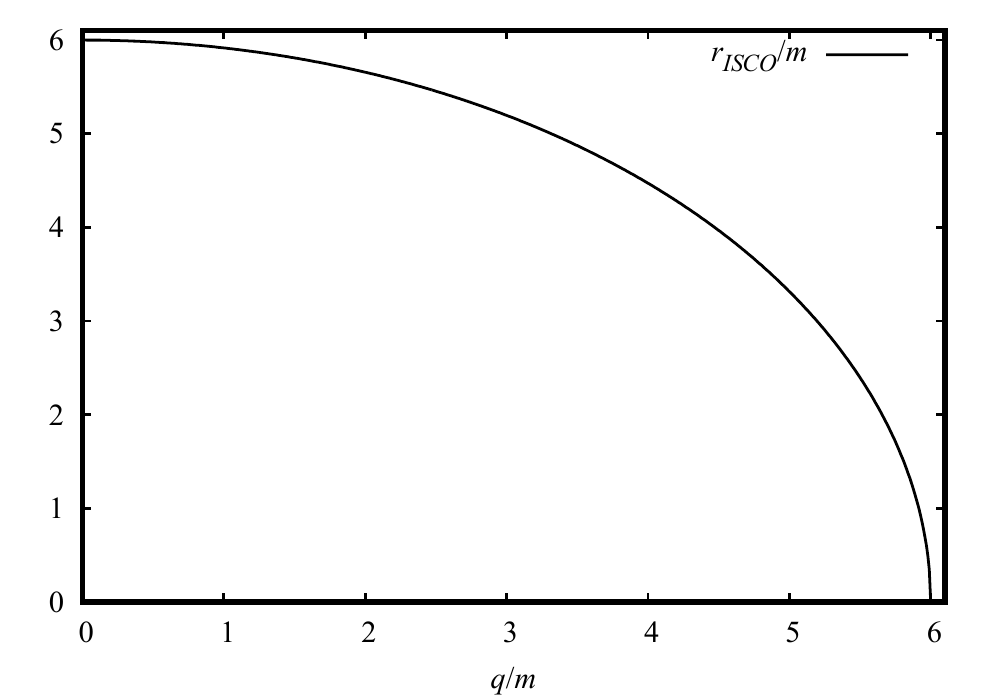}
    \caption{Radius of ISCO in the Simpson-Visser spacetime time as functions of the charge.}
    \label{fig:SV-ISCO}
\end{figure}

To identify the ISCO in the Simpson-Visser solution, set $r_1=r_3$ and solve for $l$. The result is $l_{ISCO}=2 \sqrt{3} m$, which is the same as in Schwarzschild. Thus, for $l=l_{ISCO}$, the radius is
\begin{equation}
    r_{ISCO}=\sqrt{36m^2-q^2}.
\end{equation}
In Fig. \ref{fig:SV-ISCO}, it can be observed that for $q=0$ the radius is $r_{ISCO}=6m$, which is the value to Schwarzschild, and $q=6m$ results in $r_{ISCO}\rightarrow r_5$.

\begin{figure}
    \centering
    \includegraphics[scale=0.7]{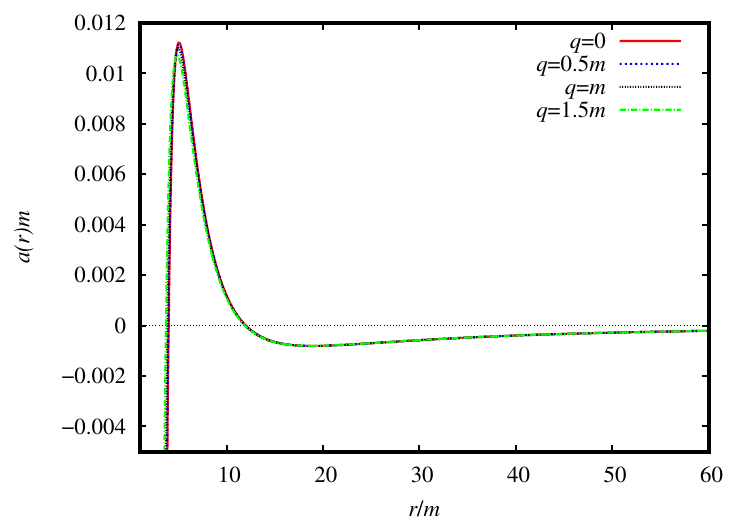}
    \includegraphics[scale=0.7]{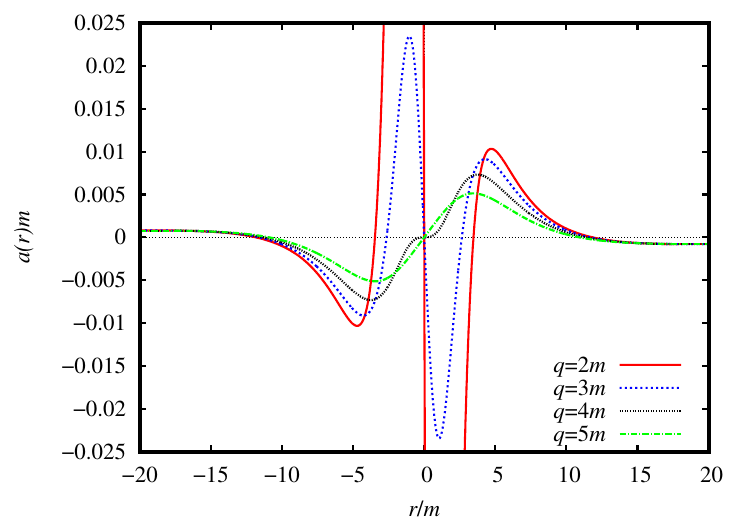}
    \includegraphics[scale=0.7]{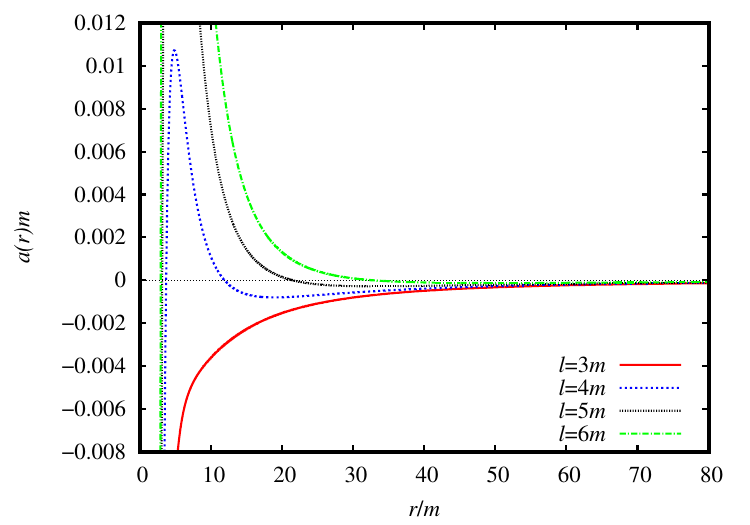}
    \includegraphics[scale=0.7]{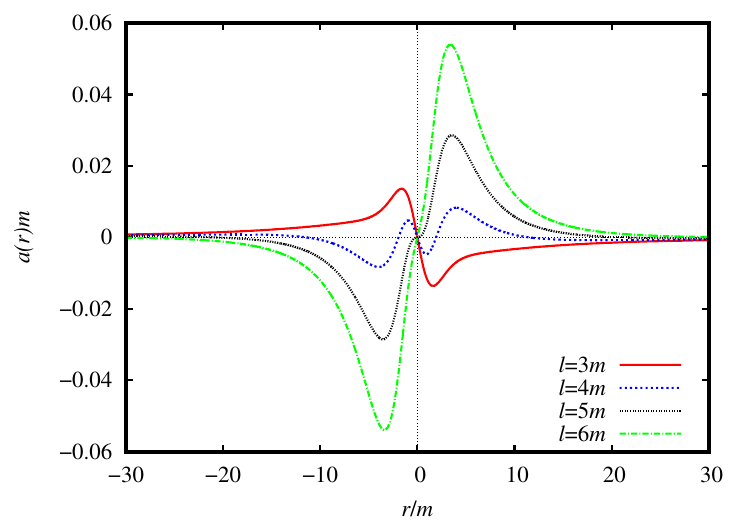}
    \caption{The radial acceleration for massive particles in the Simpson-Visser spacetime is explored for various values of charge and $l$. In (a), $l$ is held constant, $l=4m$, and the charge is varied. Since $q<2m$, there exists an event horizon, only the region $r>0$ is considered. In (b), the setup is similar to (a), but in this case, there is no event horizon, and the entire range $-\infty<r<\infty$ is considered. In (c) and (d), the charge is fixed at $q=1.5m$ and $q=3.5m$, while $l$ is adjusted.}
    \label{Fig-a-SV-massive}
\end{figure}

The radial acceleration for massive particles is
\begin{equation}
    a=-\frac{ r \left(m \left(q^2+r^2\right)-l^2 \left(\sqrt{q^2+r^2}-3
   m\right)\right)}{\left(q^2+r^2\right)^{5/2}}.\label{acc_massive}
\end{equation}
In Fig. \ref{Fig-a-SV-massive}, it is clear that, for small values of $l$, the acceleration is always attractive for $r$ positive and repulsive for $r$ negative. For other values of $l$ the sign of the acceleration change many times.

From the change of variable
\begin{equation}
    u=\frac{1}{\sqrt{r^2+q^2}},
\end{equation}
the equation \eqref{EnerConsMassive} becomes
\begin{equation}
    \left(\frac{du}{d\phi}\right)^2=\left(1-q^2u^2\right)\left[\frac{E^2-1}{l^2}+\frac{2mu}{l^2}-u^2+2mu^3\right].\label{GeoMassiveEq1}
\end{equation}
For massive particles, the impact parameter is defined as \cite{Chandrasekhar:1985kt}
\begin{equation}
    b=\frac{l}{\sqrt{E^2-1}}.
\end{equation}
The second order differential equation associated with \eqref{GeoMassiveEq1} is
\begin{equation}
    \frac{d^2u}{d\phi^2}+u=-\frac{q^2 u}{b^2}-\frac{3 m q^2 u^2}{l^2}+\frac{m}{l^2}-5 m q^2 u^4+3 m u^2+2 q^2 u^3.\label{GeoMassiveEq2}
\end{equation}
Before obtaining the trajectories, it is still necessary to verify the form of the critical impact parameter. For the critical impact parameter, the particle comes from infinity and ends up in an unstable circular orbit. At this point the radial velocity is $\dot{r}=0$. Imposing this condition in \eqref{EnerConsMassive}, the critical impact parameter is
\begin{equation}
    b_c=\frac{l}{\sqrt{E_c^2-1}}=\frac{l}{\sqrt{V_{eff}(r_{1,5})-1}}.
\end{equation}
From Fig. \ref{fig:SV-radiusmassive}, for $q<4m$, the radius of the unstable orbit is $r_1$, and for $q\geq4m$ is $r_5$. The expression for $b_c$ is not so simple for the massive case, however, for $q<4m$, $b_c$ does not depend on the charge, similar for the massless case, and for $q\geq 4m$, $b_c$ depends on the charge.

\begin{figure}
    \centering
   \subfigure[]{\includegraphics[scale=.89]{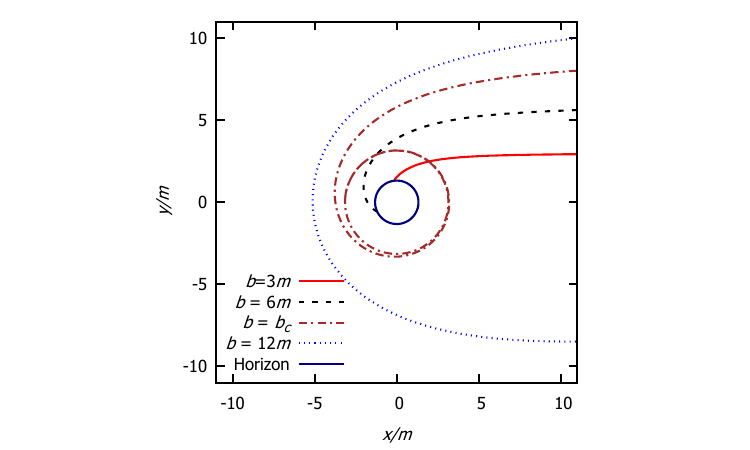}}\hspace{-4.9cm}
    \subfigure[]{\includegraphics[scale=.89]{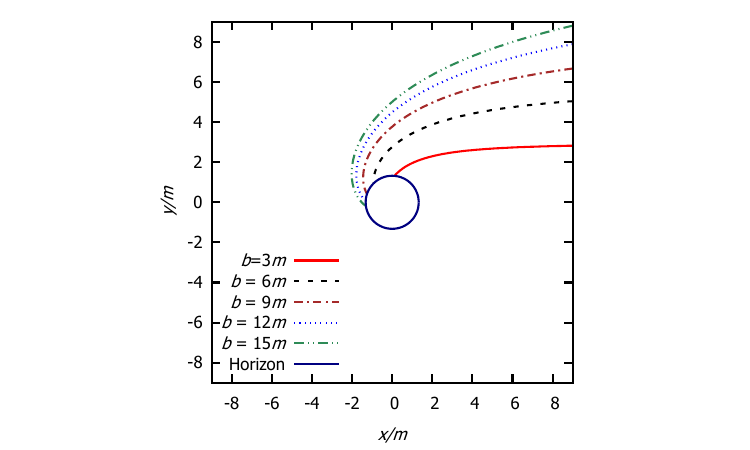}}
    \subfigure[]{\includegraphics[scale=0.89]{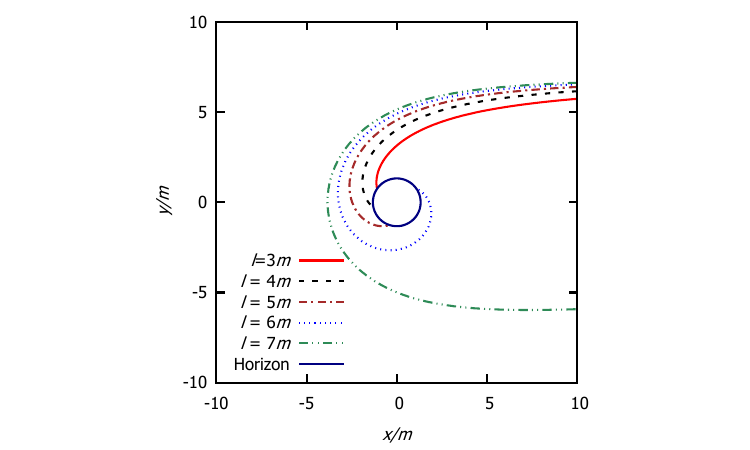}}\hspace{-4.9cm}
    \subfigure[]{\includegraphics[scale=0.89]{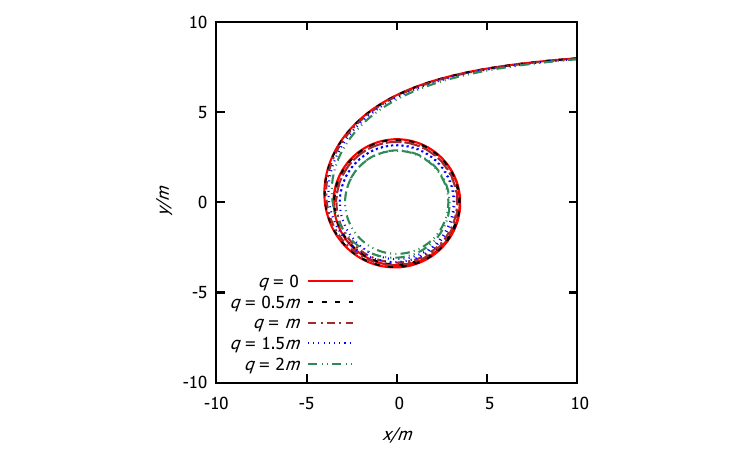}}
    \caption{Trajectories of massive particles in Simpson-Visser spacetime to different values of $q$, $l$, and $b$. In (a), the parameters are set at $q=1.5m$ and $l=5m$, while the value of $b$ is varied. In (b), $q=1.5m$ and $l=3m$ are held constant, and the value of $b$ is changed. In (c), $q=1.5m$ and $b=7m$ are fixed, and the value of $l$ is adjusted. In (d), $b$ is set to $b_c$, $l=5m$, and the value of $q$ is altered.}
    \label{fig:SV-geo-massive}
\end{figure}

In Fig. \ref{fig:SV-geo-massive}, trajectories of massive particles in Simpson-Visser spacetime are displayed. For certain values of $l$, it is possible to observe particles being scattered, absorbed, or in circular orbits, depending on the impact parameter. However, if $l$ is smaller than a specific threshold, all particles will be absorbed, regardless of the impact parameter. As the value of $l$ is increased, particles can move further before being absorbed by the black hole. When $l$ is sufficiently large, the particles are scattered. It is also noted that, while the charge does not affect the critical impact parameter, for small values of $q$, the radius of the unstable orbit decreases as the charge is increased.

\section{Causality}\label{S:causality}
Following the formalism described in the references \cite{Tomizawa:2023vir,Novello:1999pg}, it's possible to analyze the causality of photons propagating in the Simpson-Visser spacetime when considering the corrections from nonlinear electrodynamics. To do this, it's necessary consider photons in the low-energy regime.

To the effective metric, photons will obey the relation \cite{Tomizawa:2023vir,Novello:1999pg}
\begin{equation}
    g_{eff}^{\mu\nu}k_\mu k_\nu=0,\label{wave_eff}
\end{equation}
where $k_\mu=\nabla_\mu S$ is the gradient of the phase $S$, also known as the propagation vector \footnote{To more details, see ref. \cite{Novello:1999pg}}. Considering that the source is magnetically charged, equation \eqref{wave_eff} becomes
\begin{equation}
    g^{\mu\nu}k_\mu k_\nu= -\frac{2FL_{FF}}{L_F\Sigma^2}\left[\left(k_\theta\right)^2+\frac{\left(k_\varphi\right)^2}{\sin^2\theta}\right],\label{photoncausality}
\end{equation}
where $g_{\mu\nu}$ is the standard metric.

According to the notation followed in this work, we have
\begin{eqnarray}
&\mbox{for}& \quad     g^{\mu\nu}k_\mu k_\nu>0, \quad \mbox{timelike},\\
&\mbox{for}& \quad     g^{\mu\nu}k_\mu k_\nu=0, \quad \mbox{null},\\
&\mbox{for}& \quad     g^{\mu\nu}k_\mu k_\nu<0, \quad \mbox{spacelike}.
\end{eqnarray}
Photons will only follow null geodesics for radial motion. Otherwise, the trajectories will be either timelike or spacelike.

Through equation \eqref{F}, it is possible to write the functions $L_F$ and $L_{FF}$ as
\begin{equation}
    L_F=\frac{3 m}{2 \sqrt{q^2+r^2}}, \qquad L_{FF}=\frac{3 m \left(q^2+r^2\right)^{3/2}}{16 q^2}.
\end{equation}
Computing \eqref{photoncausality} to the Simpson-Visser spacetime, one obtains that
\begin{equation}
    g^{\mu\nu}k_\mu k_\nu< 0.
\end{equation}
This means that photons follow spacelike geodesics in the Simpson-Visser geometry. So that, there are faster-than-light photons \cite{Bronnikov0}.

\section{Conclusion}\label{S:conclusion}
Through the use of geodesics, it is possible to extract information about spacetime. In this study, the geodesics in a spacetime of a black bounce are examined, with the Simpson-Visser solution chosen as an example. Given the spherically symmetric nature of the solution, emphasis is placed on the equatorial plane without a loss of generality.

For massless particles, when $q\leq 3m$, the charge has a small impact on the magnitude of the effective potential. For $q<2m$, where a horizon exists, the effective potential exhibits only one extremum, which is a maximum. This implies the presence of unstable orbits for massless particles. For $2m\leq q<3m$, where no horizons are present, the effective potential features three extrema: two maxima and one minimum, indicating the existence of  stable and unstable orbits. When $q\geq3m$, there is only one extremum, which is a maximum. In the positive part of $r$, particles feel a repulsive interaction in regions beyond the maximum of the effective potential, while an attractive interaction occurs in the inner regions. In the negative part of $r$, the behavior is reversed. The critical impact parameter does not depend on the charge, for $\left|q\right|<3m$, resulting in the absorption cross-section being independent of charge.

For photons, the results are similar for the massless case. The radius of the unstable orbits is the same as for the massless case. However, the critical impact factor changes by a multiplicative term. Therefore, a photon and a massless particle must be emitted with different impact parameters to have the same orbit. The absorption cross section for photons is smaller than for massless particles so that fewer photons can be absorbed.

For massive particles, the scenario becomes more intricate. Even in cases with horizons, the number of extrema in the effective potential varies depending on the value of $l$. This results in the coexistence of stable and unstable orbits. In situations without horizons, it is possible to have three stable and two unstable orbits. The orbit radii decrease with an increase in charge, and for $q\geq 12m$, there is only one orbit at $r=0$. If $l<2\sqrt{3}m$, no orbits are present outside the black hole, and all particles are absorbed.

\begin{table}[!htpb]
	\centering
	\begin{footnotesize} 
		\setlength{\tabcolsep}{3pt}

		\begin{tabular}{|c|c|c|c|c|c|c|c|c|c|c|c|c|c|c|c|c|c|c|c|c|c|c|c|c|c|}
  \hline
		 \multicolumn{2}{|c|}{}&\multicolumn{24}{|c|}{{ \large \bf Massless particles and photons}}\\ 
   \hline 
   \multicolumn{2}{|c|}{}&
    \multicolumn{8}{|c|}{$\qquad\qquad\qquad q<2m \qquad\qquad\qquad $} &  \multicolumn{8}{|c|}{$\quad\qquad\qquad 2m\leq q <3m\quad\qquad\qquad $}& \multicolumn{8}{|c|}{$q\geq 3m$} \\ 
    \hline
        \multicolumn{2}{|c|}{$N^{\circ}$ unstable orbits}&\multicolumn{8}{|c|}{One unstable}& \multicolumn{8}{|c|}{Two unstable}& \multicolumn{8}{|c|}{One unstable} \\ 
        \hline
        \multicolumn{2}{|c|}{$N^{\circ}$ stable orbits}&\multicolumn{8}{|c|}{No stable} & \multicolumn{8}{|c|}{One stable}& \multicolumn{8}{|c|}{No stable} \\ \hline 
         \multicolumn{2}{|c|}{}&\multicolumn{24}{|c|}{The structure of the potential does not change with $l$}\\ 
         \hline
         	\multicolumn{2}{|c|}{}& \multicolumn{24}{|c|}{{\large \bf Massive particles}}\\ 
          \hline
	      \multicolumn{2}{|c|}{}&  \multicolumn{6}{|c|}{$q<2m$}&  \multicolumn{6}{|c|}{$2m\leq q <4m$}&  \multicolumn{6}{|c|}{$4m\leq q < 12m$} &  \multicolumn{6}{|c|}{$q\geq 12m$ }    \\ \hline
          \multicolumn{2}{|c|}{}& \multicolumn{3}{|c|}{$l<2\sqrt{3}m$}&  \multicolumn{3}{|c|}{$l>2\sqrt{3}m$}&  \multicolumn{3}{|c|}{$l<2\sqrt{3}m$} &  \multicolumn{3}{|c|}{$l>2\sqrt{3}m$ }&  \multicolumn{3}{|c|}{$l<2\sqrt{3}m$}&  \multicolumn{3}{|c|}{$l>2\sqrt{3}m$}&  \multicolumn{3}{|c|}{$l<2\sqrt{3}m$} &  \multicolumn{3}{|c|}{$l>2\sqrt{3}m$ }    \\
           \hline
          \multicolumn{2}{|c|}{{$N^{\circ}$ unstable orbits}}& \multicolumn{3}{|c|}{No unstable}&  \multicolumn{3}{|c|}{One unstable}&  \multicolumn{3}{|c|}{No unstable} &  \multicolumn{3}{|c|}{Two unstable }&  \multicolumn{3}{|c|}{No unstable}&  \multicolumn{3}{|c|}{One unstable}&  \multicolumn{3}{|c|}{No unstable} &  \multicolumn{3}{|c|}{No unstable}    \\
           \hline
          \multicolumn{2}{|c|}{{$N^{\circ}$ stable orbits}}& \multicolumn{3}{|c|}{No stable}&  \multicolumn{3}{|c|}{One stable}&  \multicolumn{3}{|c|}{One stable} &  \multicolumn{3}{|c|}{Three stable}&  \multicolumn{3}{|c|}{One stable}&  \multicolumn{3}{|c|}{Two stable}&  \multicolumn{3}{|c|}{One stable} &  \multicolumn{3}{|c|}{One stable}  \\ \hline 
		\end{tabular} 
	\end{footnotesize}
	\caption{Number of possible circular orbits for given values of $l$ and $q$.}
	\label{t_results}
\end{table} 

Part of the results are summarized in Table \ref{t_results}. Through this table, it is possible to determine the possible orbits for each range of black hole charge and particle angular momentum.

We also show that photons describe timelike geodesics in the Simpson-Visser geometry when the usual metric is considered and null geodesics when the effective metric is considered.

With the results obtained in this work, it is possible to study the tidal forces that particles suffer near a black hole or even explore the shadows and the optical appearance of these solutions.


\section*{Acknowledgements}
M.E.R thanks Conselho Nacional de Desenvolvimento Cient\'ifico e Tecnol\'ogico - CNPq, Brazil  for partial financial support. M.S. would like
to thank Funda\c c\~ao Cearense de Apoio ao Desenvolvimento Cient\'ifico e Tecnol\'ogico (FUNCAP) for partial financial support.



\end{document}